\documentclass[sigconf,screen]{acmart}
\copyrightyear{2026}
\acmYear{2026}
\setcopyright{cc}
\setcctype{by}
\acmConference[ASE '26]{Proceedings of the 41st IEEE/ACM International Conference on Automated Software Engineering}{October 12--16, 2026}{Munich, Germany}
\acmBooktitle{Proceedings of the 41st IEEE/ACM International Conference on Automated Software Engineering (ASE '26), October 12--16, 2026, Munich, Germany}
\acmDOI{10.1145/3832783.3837463}
\acmISBN{979-8-4007-2882-2/2026/10}
\AtBeginDocument{%
  }

\usepackage{makecell}
\usepackage{booktabs}
\usepackage{multirow}
\usepackage{xspace}
\usepackage{framed}
\usepackage{xcolor}
\usepackage{comment}
\usepackage{colortbl}
\usepackage[inline]{enumitem}
\usepackage{microtype}
\def\tool{\textbf{ReLog}\xspace}

\def\phead#1{\par\noindent{\bf\textit{#1}}\par}

\def\greybox#1{%
\begin{framed}
\setlength{\parindent}{0pt}%
\noindent\ignorespaces #1%
\end{framed}
}

\begin{document}

\title[ReLog: Execution-Aware Logging with Runtime Feedback for LLM-Oriented Debugging]{ReLog: Execution-Aware Logging with Runtime Feedback for LLM-Oriented Debugging}

\author{Xin Wang}
\authornotemark[1]
\email{xwang496@connect.hkust-gz.edu.cn}
\orcid{0009-0004-0016-9380}
\affiliation{%
  \institution{The Hong Kong University of Science and Technology (Guangzhou)}
  \city{Guangzhou}
  \country{China}
}

\author{Yang Feng}
\authornote{~Equal contribution.}
\email{1921567337@qq.com}
\orcid{0009-0000-0298-1162}
\affiliation{%
  \institution{The Hong Kong University of Science and Technology (Guangzhou)}
  \city{Guangzhou}
  \country{China}
}

\author{Xiaoqian Jiao}
\email{xjiao393@connect.hkust-gz.edu.cn}
\orcid{0009-0004-2362-7073}
\affiliation{%
  \institution{The Hong Kong University of Science and Technology (Guangzhou)}
  \city{Guangzhou}
  \country{China}
}

\author{Yang Zhang}
\email{uzhangyang@foxmail.com}
\orcid{0000-0001-8641-2660}
\affiliation{%
  \institution{Hebei University of Science and Technology}
  \city{Hebei}
  \country{China}
}

\author{Zhenhao Li}
\email{lzhenhao@yorku.ca}
\orcid{0000-0002-4909-1535}
\affiliation{%
  \institution{York University}
  \city{Toronto}
  \country{Canada}
}

\author{Zishuo Ding}
\authornote{Corresponding author.}
\orcid{0000-0002-0803-5609}
\email{zishuoding@hkust-gz.edu.cn}
\affiliation{%
  \institution{The Hong Kong University of Science and Technology (Guangzhou)}
  \city{Guangzhou}
  \country{China}
}

\renewcommand{\shortauthors}{Wang et al.}


\begin{abstract}

Logging statements are important for software debugging and maintenance. However, existing approaches to automatic logging statement generation primarily rely on static code analysis, generating statements in a single pass without considering actual program execution. Moreover, they typically evaluate generated statements by comparing them against developer-written ones, implicitly assuming that the original statements constitute an adequate gold standard. This assumption is increasingly restrictive in the LLM era: logging statements are consumed not only by human developers but also by LLMs for downstream tasks. As a result, generating logging statements solely for human developers, and evaluating them solely by similarity to developer-written statements, does not necessarily reflect their practical usefulness.

To overcome these limitations, we introduce \tool, an iterative logging generation framework guided by continuous runtime feedback. \tool leverages LLMs to generate, execute, evaluate, and iteratively refine logging statements based on runtime feedback, so that the runtime logs provide sufficient information for downstream tasks. Instead of measuring similarity to developer-written statements, we evaluate \tool through downstream tasks (i.e., defect localization and repair). We construct a new benchmark derived from Defects4J under two settings, namely direct and indirect debugging. The results show that \tool consistently outperforms all baselines across both settings. In particular, it achieves the highest bug-detection F1 score of 0.520 and successfully repairs 97 defects in the direct setting, while also obtaining the best F1 score of 0.408 in the indirect setting where source code is unavailable. Additional experiments across upstream generation models and downstream log-consuming models demonstrate the generality of the framework, while ablation studies confirm the importance of iterative refinement, compilation repair, and runtime feedback. Overall, our work reframes logging statement generation as a runtime-guided process that accounts for LLM-based downstream use, and advocates evaluating generated logging statements by their downstream utility rather than textual similarity.

\end{abstract}

\ccsdesc[500]{Software and its engineering~Software testing and debugging}
\ccsdesc[300]{Software and its engineering~Software maintenance tools}
\ccsdesc[300]{Computing methodologies~Natural language processing}

\keywords{logging statements generation, large language models, automated debugging}

\maketitle

\section{Introduction}
\label{sec:intro}

\begin{figure}[t]
\centering
\includegraphics[width=\linewidth]{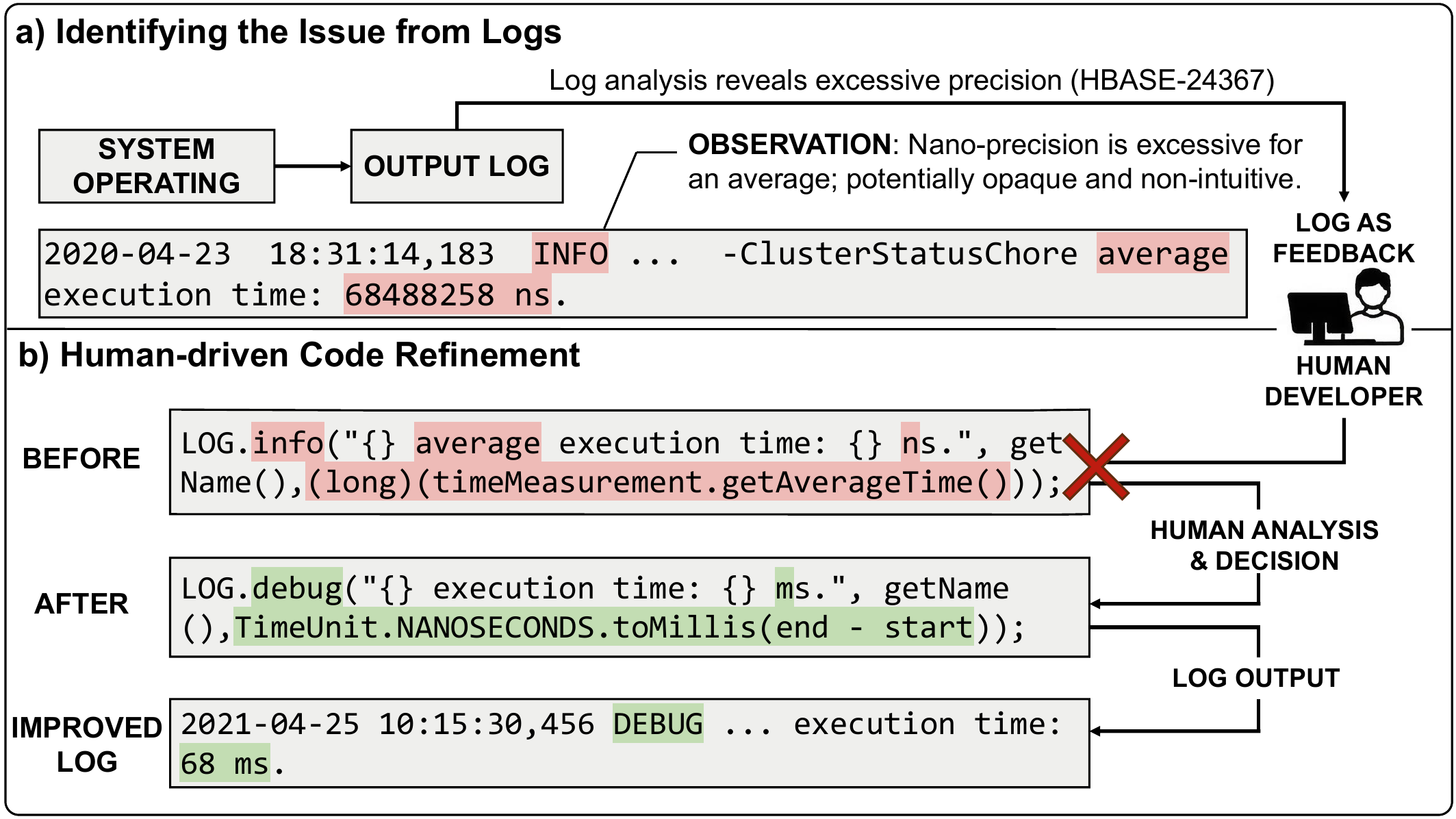}
\Description{An example showing that an Apache HBase logging statement was refined from nanosecond precision to millisecond precision and a lower severity level after runtime observation.}
\caption{The iterative refinement process of a logging statement in Apache HBase. Developers initially recorded execution time in nanoseconds but later revised the code to millisecond precision and a lower severity level based on actual runtime observations.}
\label{fig:feedbakc_refine}
\end{figure}

Software developers insert logging statements into source code to expose useful information about program execution. During execution, these statements produce runtime logs that capture detailed system behaviors and states~\cite{li2024exploring, zhong2025beyond}, thereby supporting a wide range of software engineering tasks, such as debugging, performance analysis, and system monitoring~\cite{yuan_sherlog_2010,barik_bones_2016,milani_comparative_2018,chen_an_empirical_study_leveraging_logs_2019,kim_automatic_2020,li_a_qualitative_study,he_survey_2021,li2023did}. As illustrated in Figure~\ref{fig:feedbakc_refine}, a typical logging statement consists of a severity level (e.g., info, debug), a static text template (e.g., ``\{\} execution time: {} ms.''), and dynamic variables (e.g., method name, execution time), which together record the runtime information about a program event for subsequent analysis.

Despite their importance, writing effective logging statements remains challenging for developers~\cite{chen_characterizing_2017, li_a_qualitative_study, he_survey_2021}. To alleviate this challenge, researchers have conducted a variety of studies aimed at improving logging quality and better supporting developers. Some investigate developers' logging preferences and practices~\cite{chen_characterizing_2017, hassani_studying_2018, chen_extracting_2019, li_dlfinder_2019,li_where_2020, liuWhichVariablesShould2021, foalemStudyingLoggingPractice2023, ding_temporal_2023, li_are_2023, zhong_log_updater,wang_Defects4LogBenchmarkingLLMs_2025a}, while others propose techniques for automatically suggesting or generating logging statements~\cite{zhu_learning_2015, li_where_2020, li_deeplv_2021,heng2025benchmarking, ding_logentext_2022, ding_logentext-plus_2023, li_go_2024, zhong2025beyond}. These studies largely take a developer-centric perspective, focusing on what kinds of logging statements developers prefer and how to generate statements that resemble human-written ones. However, this perspective is becoming increasingly limited in the LLM era, where logging statements are consumed not only by human developers but also by LLMs for downstream tasks such as defect localization, failure analysis, and program repair~\cite{ni_NExTTeachingLarge_2024, xu2025openrca, haque_EffectivelyLeveragingExecution_2025, kim_LogsPatchesOut_2025, ji_LeveragingLargeLanguage_2025}. Meanwhile, prior work~\cite{li_go_2024, zhong2025beyond, tan_ALBenchBenchmarkAutomatic_2025} typically evaluates generated logging statements by comparing them against developer-written ones, implicitly assuming that the original statements constitute an adequate gold standard. This assumption may also be questionable in practice, because developer-written logging statements may themselves be suboptimal~\cite{chen_characterizing_2017, li_dlfinder_2019, ding_temporal_2023, li_are_2023, wang_Defects4LogBenchmarkingLLMs_2025a}, and similarity to such statements does not necessarily reflect their practical utility for log-based downstream tasks. Therefore, we argue that automatic logging statement generation should move beyond merely imitating developer-written statements and instead aim to produce logging statements that are actually useful for downstream tasks, while also accounting for the needs of LLMs as emerging consumers of runtime logs.

In addition, existing approaches to automatic logging statement generation typically formulate the task as a static, single-pass generation problem, where the input is source code or other contextual information, such as abstract syntax trees, and the output is either a logging statement or a code snippet containing one. However, this formulation is inherently limited. In practice, logging statements are often refined iteratively based on the runtime logs they produce (cf. Section~\ref{sec:pre_study}): developers inspect the generated runtime logs, identify deficiencies in the logging statement, and then revise its content, granularity, or severity level accordingly. For example, as illustrated in Figure~\ref{fig:feedbakc_refine}, in issue HBASE-24367~\cite{HBASE-24367}, developers initially introduced a logging statement to record the average execution time of background chores in nanoseconds. However, the resulting runtime logs revealed that such fine-grained precision was neither intuitive for operators nor representative of the underlying performance characteristics. Consequently, the developers revised the logging statement to report execution time in milliseconds and adjusted the severity level to a more appropriate one. This refinement process highlights that effective logging often depends on runtime behaviors and execution states that cannot be fully inferred from static code context alone. As a result, static approaches may generate syntactically correct logging statements while still failing to capture the diagnostic information needed in practice.

To address the above limitations, we propose \tool, a runtime feedback-driven framework for iterative logging statement generation. Instead of treating logging statement generation as a static single-pass prediction task, \tool formulates it as an execution-aware refinement process. \tool is intended for offline debugging workflows, such as CI/CD failure diagnosis and local reproduction, rather than repeatedly modifying logging code on production hot paths. Given a target program and a reproducible failure, \tool first executes the original code to capture its execution outcome, and then generates initial logging statements based on both the code context and the observed runtime behavior. After inserting the generated statements, \tool compiles and executes the instrumented program, using a repair loop to resolve any compilation errors introduced by the added logging statements. It then invokes an LLM-based module to assess whether the resulting runtime logs provide sufficient observability for downstream tasks. If the diagnostic evidence is still insufficient, \tool employs a refiner module to update the logging statements based on the feedback, and repeats the process iteratively. In this way, \tool goes beyond generating logging statements that merely resemble developer-written ones, and instead aims to produce logging statements whose resulting runtime logs are useful for LLM-based downstream tasks.

To evaluate the practical utility of generated logging statements, we construct two new downstream debugging datasets from Defects4J~\cite{just_Defects4JDatabaseExisting_2014}. These datasets are designed to assess whether generated logging statements provide sufficient diagnostic information for downstream debugging under different levels of source-code availability. The first dataset, \emph{direct debugging}, contains 311 instances across 16 projects, where the LLM-based debugging agent is given both the faulty source code and the generated runtime logs for defect detection and program repair. The second dataset, \emph{indirect debugging}, contains 225 instances across 15 projects and simulates settings where the faulty source code is unavailable, requiring the agent to reason using only the generated runtime logs and caller context collected from the failing execution path.

Our evaluation shows that \tool consistently outperforms all baselines in both direct and indirect debugging. In the setting of \emph{direct debugging}, where source code is available, it achieves the highest bug detection F1 score of 0.520, exceeding the best baseline by 16.33\%, and successfully repairs 97 defects. In the setting of \emph{indirect debugging}, where debugging relies only on runtime logs and caller context, it still attains the best F1 score of 0.408, outperforming the top baseline by 16.57\%. These benefits remain consistent when varying both the LLMs used inside \tool and the downstream LLMs that consume the generated logs, indicating that the gains come mainly from the framework itself. Ablation studies further show that compilation repair, iterative refinement, and runtime feedback are essential. Finally, a qualitative case study demonstrates how the runtime feedback loop helps \tool progressively build informative evidence for exposing the root causes of defects. 


We summarize the contributions of this paper as follows:
\begin{itemize}[leftmargin=*]
    \item We propose \tool, a runtime feedback-driven framework for iterative logging statement generation. The replication package of this paper is available~\cite{replication_package}.
    \item We introduce a new evaluation methodology that assesses generated logging statements by their examining utility in downstream debugging tasks, rather than by computing similarity to developer-written statements.
    \item We construct two Defects4J-based datasets for direct and indirect debugging, enabling the evaluation of generated logging statements under different levels of source-code availability.
\end{itemize}

\noindent \textbf{Paper Organization.} The remainder of this paper is organized as follows.  Section~\ref{sec:pre_study} presents a motivating study on iterative logging refinement. Section~\ref{sec:method} introduces the design of \tool. Section~\ref{sec:exp_set} presents the experimental setup. Section~\ref{sec:rq} reports the evaluation results. Section~\ref{sec:related} reviews related work. Section~\ref{sec:threat} discusses the threats to validity, and Section~\ref{sec:conclusion} concludes the paper.

\section{Motivating Study}
\label{sec:pre_study}
\begin{figure}[t]
\centering
\includegraphics[width=\linewidth]{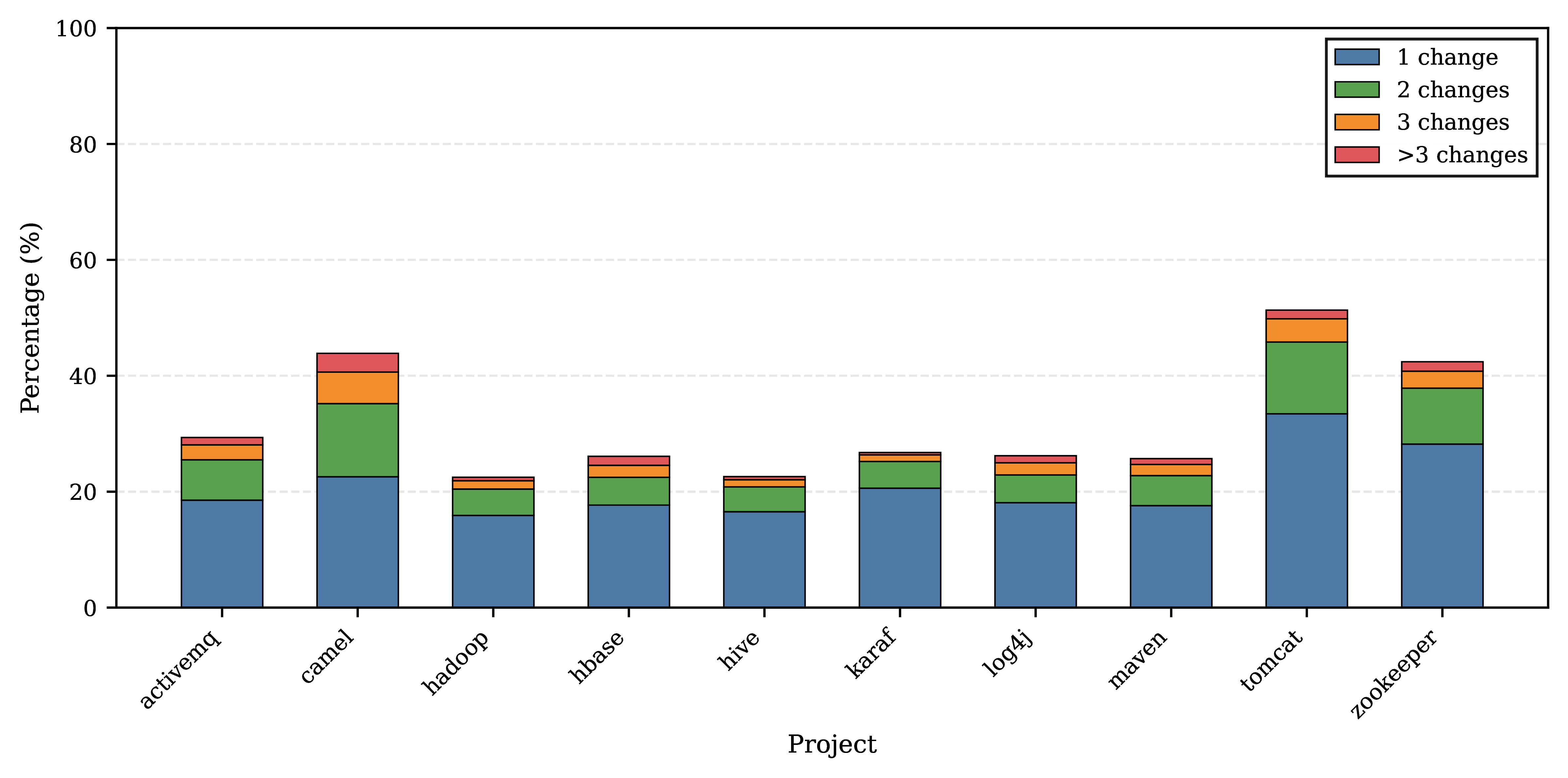}
\Description{A bar chart summarizing how often logging statements were revised across ten open source projects.}
\caption{Distribution of change frequencies for logging statements across ten open source projects. The figure displays the percentage of statements revised one to three or more times.}
\label{fig:log-change-dist}
\end{figure}

To better understand how logging statements evolve in real world software development workflows, we conduct a motivating study on their change histories. Our goal is to examine whether developers refine logging statements after their initial insertion rather than treating them as fixed instrumentation, which would provide practical motivation for iterative logging statement generation and downstream-utility-oriented evaluation.
\begin{figure*}[t]
\centering
\includegraphics[width=\textwidth]{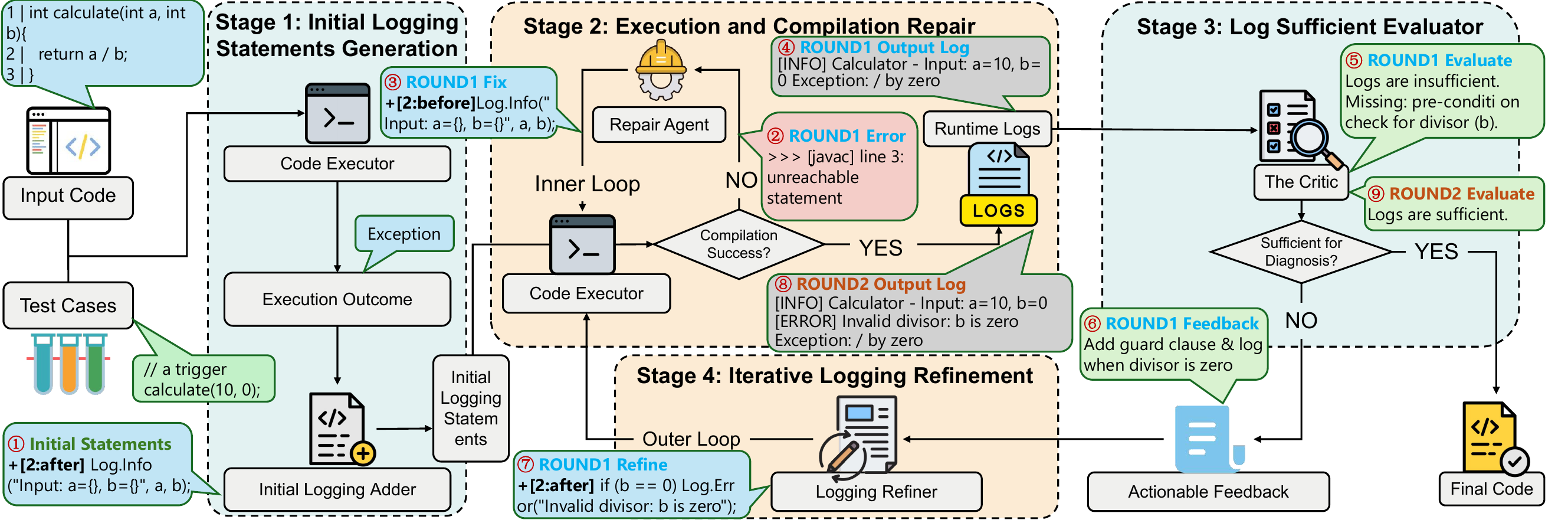}
\Description{An overview of the ReLog architecture with a two-round example of generation, execution, evaluation, and refinement.}
\caption{The overall architecture of \tool with a two-round example.}
\label{fig:architecture}
\end{figure*}
Following prior studies on logging~\cite{li_are_2023, li_go_2024, zhong2025beyond, zhong_log_updater, wang_Defects4LogBenchmarkingLLMs_2025a}, we study ten Apache projects from diverse domains: Hadoop, HBase, Hive, Tomcat, ActiveMQ, Camel, Log4j, ZooKeeper, Maven, and Karaf. For each project, we traverse its version history and identify logging statements using common logging APIs, such as \texttt{logger.debug}, and \texttt{logger.info}. We then track each logging statement across commits and count how many times it is modified. This allows to characterize how frequently logging statements are revised in practice.

Figure~\ref{fig:log-change-dist} summarizes the distribution of modification frequencies for logging statements that were changed at least once. Across all ten projects, most modified logging statements are revised only once. However, a substantial fraction is modified two or three times, and some statements are revised more than three times. Beyond this distribution, we also observe that, when considering all identified logging statements in each project, more than 20\% undergo at least one modification. In projects such as Tomcat, Camel, and ZooKeeper, this proportion even exceeds 40\%. 

These results indicate that logging statements are not always fixed after their initial insertion. Instead, developers do revise them in practice, sometimes multiple times, to better support evolving diagnostic needs. This observation suggests that treating logging statement generation as a purely static, single-pass task may not fully align with real development workflows.

It is worth noting that our measurement relies solely on committed changes recorded in version histories. In practice, developers may temporarily add, remove, or adjust logging statements during local debugging sessions before committing the final code. Such edits are typically not preserved in the repository history. Therefore, the modification frequencies reported in this study likely represent a conservative estimate, and the actual extent of logging refinement during development may be even greater~\cite{wen_EmpiricalStudyQuick_2020, chou_LearningMistakesUnderstanding_2025}. 

Overall, this motivating study provides empirical support for viewing logging statement generation as an iterative problem. It also motivates the need for approaches that account for refinement based on runtime observations, rather than relying solely on static one-pass generation.

\section{Methodology}
\label{sec:method}

\subsection{Overview}

Figure~\ref{fig:architecture} presents the overall architecture of \tool, our runtime feedback-driven framework for iterative logging statement generation. Different from existing approaches that typically formulate logging statement generation as a static and single-pass task based on source code alone, \tool treats it as an execution-aware refinement process, motivated by the observation that whether a logging statement is effective often becomes clear only after examining the runtime logs it produces during execution.

Given an input program, \tool operates as a closed-loop system with four stages. It first generates initial logging statements using both source code context and observed execution outcomes, then repairs any compilation issues introduced by the newly-added logging statements, evaluates whether the resulting runtime logs provide sufficient information for downstream tasks, and iteratively refines the logging statements when necessary. Through this iterative loop of generation, execution, evaluation, and refinement, \tool aims to produce logging statements whose resulting runtime logs are useful for downstream tasks in modern LLM-assisted software engineering workflows. We next describe these four stages in detail.

\subsection{Framework Workflow}
This subsection details the four stages of \tool's refinement loop, corresponding to the workflow illustrated in Figure~\ref{fig:architecture}.

\noindent \textbf{Stage 1: Initial Logging Statements Generation.}
Existing approaches typically generate logging statements from source code or other static contextual information alone. \tool extends this formulation by additionally incorporating runtime execution signals when available. Specifically, \tool first executes the input code snippet to obtain its initial execution outcome. If the program runs normally, the framework falls back to a source-code-based generation setting similar to prior work. If execution exposes useful diagnostic signals, such as an exception, an incorrect return value, or a timeout, \tool leverages this information to narrow the search space and focus logging generation on potentially fault-relevant parts of the code.

\tool then uses the captured execution outcome together with the input code to generate targeted initial logging statements. To prevent unintended modifications to the original logic, the framework prompts the language model to output a discrete list of logging statements paired with their specific insertion line numbers rather than regenerating the entire code snippet. The execution outcome serves as an early diagnostic signal that guides instrumentation toward code regions likely to contribute to the observed failure. For example, if a \texttt{NullPointerException} occurs at a specific line, \tool can prioritize nearby statements and relevant variables when generating logging statements. In this way, Stage~1 subsumes static logging generation as a special case while providing a more informative starting point for later stages of compilation repair, log evaluation, and iterative refinement.

\noindent \textbf{Stage 2: Compilation Repair.}
After the generated logging statements are inserted into the code, the resulting instrumented program may fail to compile. Such failures can be caused by unresolved variables, incompatible API usage, missing imports, or other inconsistencies introduced during logging insertion. A logging statement that resembles developer-written code is still of limited practical value if its insertion breaks compilation or prevents subsequent execution. Existing approaches to logging statement generation mainly focus on static generation quality, such as textual similarity to developer-written statements, while paying limited attention to whether the inserted logging statements remain compilable and executable in the target program. As we show later in our evaluation, prior approaches may generate uncompilable logging statements, which break the original program.

\tool therefore includes a compilation repair stage that uses compiler feedback to fix errors induced by the inserted logging statements. When compilation fails, the framework collects the compiler error messages and uses them to revise the generated logging code, while keeping the original program logic unchanged. Since the original program is inherently compilable, any compilation failure is strictly caused by the newly added statements. To ensure the original logic remains intact, the repair module exclusively targets and modifies the previously generated list of logging statements based on the compiler feedback. The framework then reinserts this updated list into the pristine original code. This repair loop continues until the instrumented program compiles successfully or a predefined retry limit is reached. By ensuring the executability of the inserted logging statements, this stage enables subsequent runtime evaluation and iterative refinement.

\noindent \textbf{Stage 3: Log Sufficiency Evaluation.}
After the instrumented program successfully compiles and executes, \tool collects the resulting runtime logs and evaluates whether they provide sufficient information for downstream log-based tasks. This stage is motivated by a common practice in real-world software development: developers often inspect the produced runtime logs, judge the current logging statements, and further refine them when the observed logs are still inadequate. \tool follows the same intuition, but extends this traditionally human-centered judgment process by introducing an LLM-based critic to assess whether the current logs are sufficient for downstream tasks. This design reflects the growing role of LLMs in modern software engineering workflows, as runtime logs are increasingly consumed not only by human developers but also by LLMs~\cite{xu2025openrca, haque_EffectivelyLeveragingExecution_2025, kim_LogsPatchesOut_2025, ji_LeveragingLargeLanguage_2025}.

Specifically, inspired by prior empirical studies on logging practices~\cite{yuan_sherlog_2010, barik_bones_2016, yuan_characterizing_2012, li_a_qualitative_study, hassani_studying_2018}, the critic is guided by a structured evaluation rubric to assess the current runtime logs along three predefined dimensions. The first is \textbf{traceability}, which examines whether the logs clearly expose the execution path relevant to the observed behavior. The second is \textbf{state visibility}, which checks whether key variables and intermediate states are adequately recorded. The third is \textbf{causal linkage}, which evaluates whether the logs provide enough evidence to explain why the observed behavior occurred, rather than merely showing that it occurred. Based on this structured assessment, the critic produces both a sufficiency judgment and actionable feedback for refinement. This rubric-based design is flexible and extensible: although \tool currently adopts these three dimensions, developers can incorporate additional criteria or task-specific requirements when needed, allowing the critic to adapt to different downstream scenarios beyond a fixed evaluation template.

Operationally, the critic receives the original code with line numbers, the applied logging statements, the runtime logs, and the execution outcome. It is prompted to mark the logs as sufficient only when the evidence exposes the relevant execution path, records the state needed to reason about the failure, and connects the recorded state to the observed outcome. Otherwise, it identifies the missing evidence and emits concrete refinement suggestions. The generator, fixer, critic, and refiner are implemented as separate LLM roles with separate conversations, so the sufficiency judgment is not produced by the same conversation that inserted the current statements.


Finally, if the critic determines that the current logs are sufficient, the framework terminates successfully. Otherwise, it generates feedback to guide the next refinement step. For example, the critic may identify that an important variable is updated inside a loop but never logged, and suggest inserting a logging statement at a specific location. In this way, Stage~3 serves as the key bridge between runtime log observation and iterative logging statement refinement to ensure diagnostic quality.

\noindent \textbf{Stage 4: Iterative Logging Statement Refinement.}
When the critic identifies observability gaps, \tool enters the iterative refinement stage. A logging refiner module acts as an actor that updates the current logging statements based on the structured feedback produced in Stage~3. 

Instead of regenerating logging statements from scratch, the refiner performs targeted modifications to address the identified deficiencies. Because the framework maintains the generated logging statements as a discrete list independent of the source code, the language model can directly manipulate this list. Guided by the diagnostic feedback, the refiner emits one of four logging-only actions: \textsc{Add} new statements, \textsc{Modify} the variables or message content of existing statements, \textsc{Remove} redundant statements, or \textsc{Relocate} statements to more informative execution points. These actions may add missing variable states, adjust logging position to better capture dynamic control flow, or remove redundant information that introduces noise. The refiner is explicitly constrained not to change the original program logic; the updated list is reinserted into the pristine original code before the next compilation and execution. This design allows \tool to preserve useful existing instrumentation while incrementally improving the parts that limit downstream log utility.

Once the logging statements are updated, the revised code is fed back into Stage~2 for compilation repair and execution, and the resulting runtime logs are reevaluated in Stage~3. The loop continues until the logs are judged sufficient for downstream tasks or a predefined maximum number of iterations is reached (five in our experiments). In this way, \tool progressively improves the diagnostic quality of the generated logging statements through a closed-loop refinement process.




\section{Experiment Setup}
\label{sec:exp_set}
\subsection{Downstream Evaluation Task}
Prior work~\cite{li_go_2024, zhong2025beyond, tan_ALBenchBenchmarkAutomatic_2025} on logging statement generation is typically evaluated against developer-written statements using static matching-based metrics, such as position and level accuracy, and textual similarity. However, as discussed in the \textbf{Introduction}, this evaluation  has limitations: developer-written logging statements may themselves be suboptimal, and static metrics do not directly capture whether the resulting runtime logs are useful for downstream tasks.  To address this limitation, we propose to evaluate \tool through downstream tasks to assess whether the generated logging statements provide useful diagnostic information for practical debugging scenarios.

Specifically, we choose debugging tasks as the evaluation target, since it is a primary application of runtime logs~\cite{ni_NExTTeachingLarge_2024, chen_an_empirical_study_leveraging_logs_2019, chen_PathideaImprovingInformation_2022, haque_EffectivelyLeveragingExecution_2025, kim_LogsPatchesOut_2025} and serves as a practical testbed for assessing whether the generated logging statements provide useful diagnostic information. In particular, we consider two settings that reflect different levels of source-code availability: \emph{direct debugging}, where the faulty code is accessible, and \emph{indirect debugging}, where diagnosis must rely on runtime logs and calling context alone.

\subsubsection{Direct Debugging}

In this setting, the debugging agent has access to both the faulty source code and the generated runtime logs. This setup aligns with common development scenarios, where engineers can inspect the code while using logs to understand the failure.

This task consists of two distinct subtasks:
\begin{enumerate*}[label=(\roman*)]
\item \textbf{Defect Localization.} Given the code and the corresponding runtime logs, the agent determines whether a defect exists and produces a textual description of the identified issue. A prediction is considered a \emph{True Positive} when the generated defect description matches the ground truth, and  Precision, Recall, and F1 Score are used to evaluate localization effectiveness.
\item \textbf{Program Repair.} If a defect is detected, the agent is prompted to repair the faulty code using the generated defect description together with the available runtime information. The final output is a candidate patch. A repair is considered successful if the generated patch is identical or semantically equivalent to the ground-truth fix.

\end{enumerate*} 


\subsubsection{Indirect Debugging}
In this setting, the faulty method itself is inaccessible to the downstream debugging agent. The debugging process therefore relies on runtime logs and caller-side context observed on the failing execution path. This setup reflects operational environments where developers may not have direct access to the faulty source file, but can still inspect surrounding caller context or failure-triggering traces.

Accordingly, this setting focuses only on \textbf{defect localization}. Given the runtime logs produced by the instrumented program and the available calling context, the agent determines whether a defect exists and generates a textual description of the issue. This scenario evaluates whether the generated logs alone provide sufficient information to diagnose failures when the source code is hidden. Similarly, we use Precision, Recall, and F1 Score as the evaluation metric.



\subsubsection{Debugging Agents}

LLM-based agents are increasingly used to assist developers in software engineering tasks~\cite{jimenez_SWEbenchCanLanguage_2024, xu2025openrca}. As a result, it is increasingly important to provide runtime logs that are useful for LLM-based tasks. Following this trend, we implement lightweight LLM-based debugging agents to perform the above-mentioned tasks in our evaluation. For localization, the agent takes the available code context and runtime logs as input, and outputs a defect decision together with a textual description of the issue. For repair, the agent takes the faulty code and the generated defect description as input, and outputs a candidate patch. Our purpose is not to propose a state-of-the-art agent for defect localization or repair, but to examine whether the generated runtime logs provide useful information for LLM-based debugging.

To evaluate localization, two authors independently compare each predicted defect description with the ground-truth defect and mark it correct when it identifies the same faulty behavior and causal mechanism, rather than requiring exact wording. To evaluate repair, they mark a patch correct when it is textually identical to the ground-truth fix or implements the same behavior. Disagreements are resolved by a third author.

\subsection{Dataset Construction}
To support the downstream evaluation, we construct our datasets from the widely used Defects4J benchmark~\cite{just_Defects4JDatabaseExisting_2014}, as existing resources do not directly provide such settings. Based on different levels of source-code availability, we derive two datasets: one for direct debugging and the other for indirect debugging. Note that our goal is not to advance the state of the art on Defects4J itself, but to use Defects4J-based downstream debugging tasks as a practical testbed for evaluating whether generated logging statements provide useful diagnostic information.




\noindent \textbf{Direct Debugging Dataset:} Each sample is centered on a single faulty method. We retain reproducible Defects4J cases whose faulty method can be identified by comparing buggy and fixed versions, whose buggy and fixed method bodies are extractable, and whose triggering test reliably reproduces the failure. The resulting dataset contains 311 samples from 16 projects.

\noindent \textbf{Indirect Debugging Dataset:} For each sample, we additionally extract the direct caller methods that appear on the actual invocation path leading to the failure. We obtain these callers by instrumenting test executions and recording method invocation chains, while retaining only project source methods with resolvable bodies. The downstream agent cannot access the faulty method, but can access the extracted caller context. The resulting dataset contains 225 samples from 15 projects, each augmented with at least one caller method resolved at runtime.




\subsection{Baselines}

We evaluate our approach against several representative baselines that span diverse automated logging generation paradigms, ranging from traditional deep neural networks to modern LLMs.

    
    
    
    

\begin{itemize}[leftmargin=*]
    \item \textbf{SCLogger}~\cite{li_go_2024} generates logging statements by leveraging static cross method contexts and variable type information to determine logging positions, severity levels, and message content.
    
    \item \textbf{UniLog}~\cite{xu_UniLogAutomaticLogging_2024} relies on in context learning with demonstration examples to generate logging statements.
    
    \item \textbf{LANCE}~\cite{mastropaolo_UsingDeepLearning_2022} utilizes deep learning to predict the insertion position and generate a single textual logging statement.
    
    \item \textbf{LANCE2}~\cite{mastropaolo_LogStatementsGeneration_2023} extends LANCE by combining deep learning and information retrieval to determine logging necessity and support multiple statement injections.
    
    \item \textbf{FastLog}~\cite{xie_FastLogEndtoEndMethod_2024} improves generation efficiency by predicting token level insertion positions to construct complete logging statements during the generation process.
\end{itemize}

\begin{table*}[htbp]
\centering
\caption{Comparison with Baseline Logging Approaches on Direct and Indirect Debugging Tasks}
\label{tab:baseline_comparison}

\resizebox{\textwidth}{!}{
\setlength{\tabcolsep}{10pt}
\renewcommand{\arraystretch}{0.91}
\begin{tabular}{l | c c c c c c c c}
\toprule
\multirow{2}{*}{Method} & \multicolumn{8}{c}{Direct Debugging} \\
\cmidrule(lr){2-9}
& Compilation Failures & Detected Defects & True Positives & Precision & Recall & F1 Score & Successful Repairs & Avg. Logs \\
\midrule
\rowcolor{gray!20}
\tool & \textbf{0} & \textbf{300} & \textbf{159} & \textbf{0.530} & \textbf{0.511} & \textbf{0.520} & \textbf{97} & 5.52 \\
\tool$_{1\text{-}log}$ & 0 & 276 & 142 & 0.514 & 0.457 & 0.484 & 84 & 1.00 \\
UniLog & 30 & 270 & 130 & 0.481 & 0.418 & 0.447 & 63 & 0.97 \\
GoStatic & 18 & 288 & 129 & 0.448 & 0.415 & 0.431 & 78 & 7.63 \\
Without Log & 0 & 283 & 124 & 0.438 & 0.399 & 0.418 & 72 & 0 \\
LANCE & 74 & 225 & 77 & 0.342 & 0.248 & 0.288 & 38 & 0.58 \\
FastLog & 142 & 162 & 65 & 0.401 & 0.209 & 0.275 & 34 & 0.88 \\
LANCE2 & 171 & 135 & 38 & 0.281 & 0.122 & 0.170 & 22 & 0.76 \\
\midrule
\multirow{2}{*}{Method} & \multicolumn{8}{c}{Indirect Debugging} \\
\cmidrule(lr){2-9}
& Compilation Failures & Detected Defects & True Positives & Precision & Recall & F1 Score & Avg. Logs per Caller \\
\midrule
\rowcolor{gray!20}
\tool & \textbf{2} & \textbf{142} & \textbf{75} & \textbf{0.528} & \textbf{0.333} & \textbf{0.408} & 6.21 \\
GoStatic & 78 & 112 & 59 & 0.527 & 0.262 & 0.350 & 9.76 \\
\tool$_{1\text{-}log}$ & 29 & 100 & 57 & 0.570 & 0.253 & 0.350 & 1.00 \\
UniLog & 12 & 64 & 32 & 0.500 & 0.142 & 0.221 & 0.94 \\
LANCE & 56 & 52 & 21 & 0.404 & 0.093 & 0.151 & 0.47 \\
LANCE2 & 184 & 27 & 12 & 0.444 & 0.053 & 0.095 & 0.66 \\
FastLog & 134 & 42 & 12 & 0.286 & 0.053 & 0.089 & 0.54 \\
\bottomrule
\end{tabular}
}
\end{table*}



\section{Research Questions}
\label{sec:rq}

In this section, we aim to answer the following research questions:

\subsection*{RQ1: How Well Does \tool Support Downstream Debugging Tasks?}

\noindent \textbf{\textit{Motivation.}}
Existing approaches to logging statement generation are typically static and evaluated using surface-level matching metrics. In contrast, \tool introduces runtime feedback and iterative refinement. The key question is whether this design improves the practical usefulness of generated logging statements. Therefore, we investigate whether \tool more effectively supports downstream debugging tasks, including defect localization and program repair.

\noindent \textbf{\textit{Approach.}}
We conduct our experiments using the direct debugging and indirect debugging datasets. For each defective code snippet, we apply \tool and all selected baselines to generate logging statements. To ensure a fair comparison, \tool and all baselines are based on DeepSeek-V3 because of its balance between cost and efficiency. We attempt to compile and execute the updated code to collect the generated runtime logs. We record any compilation failure as an unsuccessful attempt. Finally, we evaluate the successfully collected logs through downstream debugging tasks.



\noindent \textbf{\textit{Results.}}
\textbf{\tool achieves the highest overall debugging performance in the direct debugging setting.}
As shown in Table~\ref{tab:baseline_comparison}, \tool attains the highest F1 score for defect localization at \textbf{0.520}. This performance surpasses UniLog with an F1 score of 0.447, GoStatic at 0.431, and the baseline without logging at 0.418. This strong performance stems from both a higher precision of 0.530 and a recall of 0.511. These metrics indicate that \tool detects more true defects while simultaneously producing fewer incorrect detections. In terms of absolute numbers, \tool correctly identifies 159 out of 311 defects. Meanwhile, the strongest baseline, UniLog, detects 130 defects. Furthermore, \tool achieves the highest repair performance by successfully fixing \textbf{97 defects}, which accounts for 31.19\% of the total cases. This result surpasses GoStatic with 78 repairs and UniLog with 63 repairs. Compared to earlier deep learning approaches such as LANCE with 38 repairs and FastLog with 34 repairs, \tool increases the repair success rate by more than a factor of two. These results demonstrate that logging refinement guided by runtime feedback produces statements that effectively assist downstream debugging and repair tasks.

\textbf{\tool also consistently outperforms all baselines in the indirect debugging setting.}
When the source code is unavailable and debugging relies primarily on runtime logs, \tool still achieves the highest localization performance with an F1 score of \textbf{0.408}. This result exceeds the strongest baseline, GoStatic, which achieves an F1 score of 0.350. In absolute terms, \tool correctly detects 75 out of 225 defects. In comparison, GoStatic detects 59 defects and UniLog detects 32 defects. The performance gap becomes even larger when compared to deep learning approaches such as LANCE with an F1 score of 0.151 and LANCE2 with an F1 score of 0.095. These findings indicate that the iterative refinement process produces statements capturing informative runtime behaviors to enable reliable debugging even when the underlying source code remains inaccessible during the diagnosis.

\textbf{\tool maintains a consistently lower compilation failure rate than most baselines.}
In the direct debugging scenario, \tool introduces zero compilation failures. Conversely, several baselines frequently generate uncompilable code. For example, FastLog produces 142 failures and LANCE2 produces 171 failures. A similar trend appears in the indirect debugging setting. Here, \tool results in only two compilation failures, compared to 78 failures for GoStatic and 184 failures for LANCE2. This stability highlights how the integrated repair module successfully maintains executable logging code throughout the entire refinement process.

\textbf{Generated logs provide additional value beyond source code alone.}
The direct debugging setting gives the downstream agent access to the faulty source code, so the ``Without Log'' baseline helps isolate the contribution of generated runtime logs. Removing generated logs reduces F1 from 0.520 to 0.418 and successful repairs from 97 to 72. This shows that \tool's logs provide diagnostic evidence beyond the code itself. The gap is even clearer in the indirect setting, where the agent cannot inspect the faulty method and must rely on logs plus caller context.

\textbf{\tool overcomes baseline limitations by generating sufficient diagnostic logging statements.}
Evaluation results show that deep learning baselines like FastLog, LANCE, and LANCE2 average fewer than one logging statement per method. This restriction stems from their fundamental task formulation, as these approaches are designed for a one shot generation paradigm. Although LANCE2 incorporates multiple generation, its objective remains focused on optimizing individual predictions instead of producing comprehensive diagnostic traces. Likewise, UniLog relies on in context learning with demonstrations derived from the same single insertion paradigm, leading to a comparable bias of 0.97 logs per method. These designs inherently restrict the capacity to capture complex execution states, since a single statement rarely exposes the necessary control flow and intermediate variables for debugging. In contrast, \tool adopts an unconstrained paradigm guided by runtime feedback, enabling the generation and iterative refinement of multiple logging statements. Consequently, \tool produces highly informative diagnostic traces, averaging 5.52 logs per method in the direct setting and 6.21 per caller in the indirect setting. Furthermore, \tool remains more concise than static analysis baselines like GoStatic, which generate 7.63 and 9.76 logs respectively. These findings demonstrate that \tool transcends the constraints of single statement generation, ensuring sufficient diagnostic observability without excessive verbosity.

\textbf{\tool remains effective under a one-log budget.}
Average log count is a descriptive overhead statistic rather than a direct measure of log sufficiency, since one statement may record multiple variables. To address logging-budget fairness, we additionally evaluate \tool$_{1\text{-}log}$, a constrained variant that permits only one generated logging statement. In direct debugging, this variant achieves an F1 score of 0.484, higher than all non-\tool baselines, while using the same average log budget as single-insertion methods. In indirect debugging, it reaches an F1 score of 0.350, matching GoStatic and exceeding UniLog, LANCE, LANCE2, and FastLog. These results show that \tool's gains are not solely explained by inserting more statements; adaptive multi-log refinement further improves performance when the debugging task requires richer observability.

\greybox{
\textbf{RQ1 Summary:}
\tool consistently outperforms existing baseline approaches across both direct and indirect debugging scenarios. By successfully improving defect localization and program repair rates, \tool demonstrates strong robustness whether the source code is accessible or completely unavailable. These outcomes confirm that iterative refinement guided by execution feedback produces high quality logging statements, providing benefits for downstream debugging tasks.
}


\begin{table*}[htbp]
\centering
\caption{Impact of Different LLMs on \tool Performance}
\label{tab:different_llms}

\resizebox{\textwidth}{!}{
\setlength{\tabcolsep}{10pt}
\renewcommand{\arraystretch}{0.90}
\begin{tabular}{l | c c c c c c c c}
\toprule
\multirow{2}{*}{Model} & \multicolumn{8}{c}{Direct Debugging} \\
\cmidrule(lr){2-9}
& Compilation Failures & Detected Defects & True Positives & Precision & Recall & F1 Score & Successful Repairs & Avg. Logs \\
\midrule
DeepSeek-V3 & 0 & 300 & 159 & 0.530 & 0.511 & 0.520 & 97 & 5.52 \\
Qwen3-Coder-30B & 3 & 299 & 158 & 0.528 & 0.508 & 0.518 & 84 & 7.50 \\
GLM-4.7 & 0 & 299 & 168 & 0.562 & 0.540 & 0.551 & 92 & 4.41 \\
GPT-5-mini & 2 & 297 & 170 & 0.572 & 0.547 & \textbf{0.559} & 90 & 6.43 \\
\midrule
\multirow{2}{*}{Model} & \multicolumn{8}{c}{Indirect Debugging} \\
\cmidrule(lr){2-9}
& Compilation Failures & Detected Defects & True Positives & Precision & Recall & F1 Score & Successful Repairs & Avg. Logs per Caller \\
\midrule
DeepSeek-V3 & 2 & 142 & 75 & 0.528 & 0.333 & 0.408 & -- & 6.21 \\
Qwen3-Coder-30B & 12 & 113 & 64 & 0.566 & 0.284 & 0.378 & -- & 5.60 \\
GLM-4.7 & 5 & 116 & 65 & 0.560 & 0.289 & 0.382 & -- & 3.86 \\
GPT-5-mini & 8 & 148 & 80 & 0.541 & 0.356 & \textbf{0.430} & -- & 8.06 \\
\bottomrule
\end{tabular}
}
\end{table*}

\subsection*{RQ2: How Does \tool Perform Across Different Large Language Models?}

\noindent \textbf{\textit{Motivation.}}
Our framework utilizes LLMs for the generation, evaluation, and refinement of logging statements. However, these models vary substantially in their parameter size, training data, reasoning capacity, and ability to follow instructions. It remains unclear whether the effectiveness of \tool depends on a specific model or whether the proposed refinement mechanism based on runtime feedback generalizes across different models. Evaluating \tool with multiple models allows us to assess its robustness and helps determine whether the observed performance benefits stem from the iterative refinement design rather than the inherent capabilities of a particular model.



\noindent \textbf{Approach.}
We evaluate \tool using several representative LLMs with varying scales and architectures across both datasets. To ensure consistent evaluation, we maintain identical experimental settings for all models, including the limits on refinement iterations and compilation repair strategies. First, we vary the upstream LLM used by the \tool generator, fixer, critic, and refiner while keeping the downstream debugging agent fixed. Second, to test whether generated logs remain useful to different log consumers, we reuse each method's generated logs and evaluate downstream localization with GPT-5-mini, Qwen3-Coder-30B, and GLM-4.7. We assess whether the iterative refinement mechanism consistently enhances downstream debugging performance regardless of the underlying generation or consumption model.

\noindent \textbf{\textit{Results.}}
\textbf{\tool demonstrates strong robustness across different language models in the direct debugging setting.}
As shown in Table~\ref{tab:different_llms}, all evaluated models achieve comparable performance when integrated into \tool. This indicates that the effectiveness of the framework is not strictly tied to a specific model. Among the tested LLMs, GPT-5-mini achieves the highest defect localization performance with an F1 score of \textbf{0.559}. GLM-4.7 follows closely with an F1 score of 0.551. DeepSeek-V3 and Qwen3-Coder-30B achieve scores of 0.520 and 0.518 respectively. These results show that the localization performance remains consistently high across different models, with the differences falling within a narrow margin. Furthermore, all models achieve strong repair performance. GPT-5-mini successfully repairs 90 bugs, and GLM-4.7 repairs 92 bugs. This consistent success demonstrates that the iterative refinement mechanism effectively guides the debugging process regardless of the underlying language model.

\textbf{\tool maintains consistent performance across models in the indirect debugging setting.}
GPT-5-mini again achieves the highest localization performance with an F1 score of \textbf{0.430}. DeepSeek-V3 follows with a score of 0.408, while GLM-4.7 and Qwen3-Coder-30B achieve 0.382 and 0.378 respectively. In terms of absolute localization counts, GPT-5-mini correctly identifies 80 defects out of 225 total cases, while DeepSeek-V3 detects 75 defects. Although the models differ in their architecture and scale, the overall performance gap remains moderate. This observation suggests that the debugging effectiveness primarily originates from the framework design rather than the specific capabilities of the underlying model.

\textbf{The performance variations across models are smaller than the overall benefits provided by the \tool framework.}
When compared against the baseline methods reported in RQ1, all tested LLM variants within \tool consistently outperform existing logging generation approaches. For instance, even the model with the lowest performance in the direct debugging setting, Qwen3-Coder-30B with an F1 score of 0.518, still surpasses the strongest baseline shown in Table~\ref{tab:baseline_comparison}. This finding indicates that the refinement mechanism guided by runtime feedback plays a more critical role in assisting debugging tasks than the selection of LLMs.

\textbf{Different language models exhibit varying logging generation behaviors while maintaining stable debugging performance.}
For example, Qwen3-Coder-30B generates the highest number of logs on average, producing 7.50 statements per method in the direct debugging scenario. In contrast, GLM-4.7 generates fewer logs with an average of 4.41 statements. Despite these distinct differences in logging quantity, all models achieve similar levels of debugging effectiveness. This outcome suggests that the iterative refinement process successfully adapts the generated logs to the specific requirements of the debugging task. It ensures that the necessary runtime information is captured regardless of the specific generation style of the chosen model.

\begin{table}[t]
\centering
\caption{F1 Scores with Different Downstream LLM Log Consumers under Direct and Indirect Debugging}
\label{tab:cross_downstream_llms}

\resizebox{\columnwidth}{!}{
\setlength{\tabcolsep}{4pt}
\renewcommand{\arraystretch}{0.90}
\begin{tabular}{l | c c c c c c}
\toprule
\multirow{2}{*}{Downstream LLM} & \multicolumn{6}{c}{Direct Debugging} \\
\cmidrule(lr){2-7}
& \tool & UniLog & GoStatic & LANCE & FastLog & LANCE2 \\
\midrule
DeepSeek-V3 & \textbf{0.520} & 0.447 & 0.431 & 0.288 & 0.275 & 0.170 \\
GPT-5-mini & \textbf{0.678} & 0.593 & 0.570 & 0.462 & 0.349 & 0.302 \\
Qwen3-Coder-30B & \textbf{0.421} & 0.305 & 0.369 & 0.218 & 0.179 & 0.145 \\
GLM-4.7 & \textbf{0.552} & 0.488 & 0.454 & 0.331 & 0.291 & 0.188 \\
\midrule
\multirow{2}{*}{Downstream LLM} & \multicolumn{6}{c}{Indirect Debugging} \\
\cmidrule(lr){2-7}
& \tool & UniLog & GoStatic & LANCE & FastLog & LANCE2 \\
\midrule
DeepSeek-V3 & \textbf{0.408} & 0.221 & 0.350 & 0.151 & 0.089 & 0.095 \\
GPT-5-mini & \textbf{0.459} & 0.363 & 0.366 & 0.273 & 0.213 & 0.150 \\
Qwen3-Coder-30B & \textbf{0.369} & 0.147 & 0.221 & 0.109 & 0.084 & 0.033 \\
GLM-4.7 & \textbf{0.462} & 0.253 & 0.370 & 0.174 & 0.110 & 0.083 \\
\bottomrule
\end{tabular}
}
\end{table}

\textbf{\tool-generated logs remain useful when consumed by different downstream LLMs.}
Table~\ref{tab:cross_downstream_llms} reports F1 scores when the downstream log consumer is changed while the generated logs are reused. Across all four downstream LLMs and both debugging settings, \tool achieves the best F1 score among all logging methods. With DeepSeek-V3 as the downstream consumer, \tool reaches 0.520 F1 in direct debugging and 0.408 F1 in indirect debugging. With GPT-5-mini, \tool reaches 0.678 and 0.459. With the locally hosted Qwen3-Coder-30B, \tool still obtains the highest F1 scores of 0.421 and 0.369. With GLM-4.7, \tool again ranks first with 0.552 and 0.462. This cross-consumer analysis reduces the risk that \tool merely tailors logs to a single downstream agent and supports the claim that runtime-refined logs encode generally useful diagnostic evidence.

\greybox{
\textbf{RQ2 Summary:}
\tool consistently achieves strong debugging performance across different upstream generation models and downstream log-consuming models. Although performance varies across models, \tool remains the top-performing logging method in every evaluated cross-consumer setting. This consistency confirms that the effectiveness of \tool primarily stems from its iterative refinement mechanism guided by runtime feedback, rather than a strict reliance on any particular language model.
}

\subsection*{RQ3: What Is the Impact of Key Components in \tool?}

\begin{table*}[htbp]
\centering
\caption{Ablation Study of \tool Components}
\label{tab:ablation}

\resizebox{\textwidth}{!}{
\setlength{\tabcolsep}{10pt}
\renewcommand{\arraystretch}{0.91}
\begin{tabular}{l | c c c c c c c c}
\toprule
\multirow{2}{*}{Variant} & \multicolumn{8}{c}{Direct Debugging} \\
\cmidrule(lr){2-9}
& Compilation Failures & Detected Defects & True Positives & Precision & Recall & F1 Score & Successful Repairs & Avg. Logs \\
\midrule
\tool & 0 & 300 & 159 & 0.530 & 0.511 & \textbf{0.520} & 97 & 5.52 \\
\tool w/o Fixer & 94 & 208 & 114 & 0.548 & 0.366 & 0.439 & 75 & 4.79 \\
\tool w/o Runtime Feedback & 17 & 287 & 138 & 0.481 & 0.443 & 0.461 & 76 & 1.79 \\
\tool w/o Refine & 1 & 297 & 118 & 0.397 & 0.379 & 0.388 & 81 & 4.50 \\
\midrule
\multirow{2}{*}{Variant} & \multicolumn{8}{c}{Indirect Debugging} \\
\cmidrule(lr){2-9}
& Compilation Failures & Detected Defects & True Positives & Precision & Recall & F1 Score & Avg. Logs per Caller \\
\midrule
\tool & 2 & 142 & 75 & 0.528 & 0.333 & \textbf{0.408} & 6.21 \\
\tool w/o Fixer & 106 & 102 & 58 & 0.569 & 0.258 & 0.355 & 6.37 \\
\tool w/o Runtime Feedback & 29 & 122 & 62 & 0.554 & 0.276 & 0.368 & 12.50 \\
\tool w/o Refine & 14 & 116 & 62 & 0.534 & 0.276 & 0.364 & 5.41 \\
\bottomrule
\end{tabular}
}
\end{table*}

\noindent \textbf{\textit{Motivation.}}
\tool integrates three core mechanisms: a compilation repair module, iterative refinement, and runtime feedback. To validate the overall architecture of our framework, we isolate and evaluate the individual contribution of each element. Specifically, we investigate whether the observed performance benefits originate primarily from successfully resolving compilation errors, from multi-round source-based refinement, or from continuously enhancing the logging statements through actual execution results. This ablation analysis justifies our specific design choices and confirms the necessity of these mechanisms to achieve optimal performance.

\noindent \textbf{\textit{Approach.}}
We evaluate the individual contribution of each core component by testing three ablated variants of \tool. We maintain identical experimental settings across all evaluations. The first variant completely disables the compilation repair module. Consequently, the generation process terminates immediately upon encountering any compilation error. The second variant disables runtime feedback while preserving the same multi-round source-based refinement budget, so the model can revise logs but cannot use actual runtime logs as refinement evidence. The third variant removes the iterative refinement loop, limiting the framework to a single static generation pass. Comparing these results against the full framework quantifies the specific impact of each component.





\noindent \textbf{\textit{Results.}}
\textbf{Compilation repair, runtime feedback, and iterative refinement each contribute measurably to \tool.}
As shown in Table~\ref{tab:ablation}, disabling any of these mechanisms leads to a noticeable degradation in debugging performance across both evaluation settings. In the direct debugging scenario, the full version of \tool achieves an F1 score of \textbf{0.520}. It correctly detects 159 bugs and successfully repairs 97 defects. When we remove the compilation fixer, the F1 score drops to 0.439, and the number of successful repairs decreases to 75. Disabling runtime feedback reduces the F1 score to 0.461 and successful repairs to 76. Removing the refinement mechanism reduces the F1 score further to 0.388 and brings the successful repairs down to 81. These consistent drops indicate that all three mechanisms play important roles in assisting downstream debugging tasks.

\textbf{The compilation repair module is essential for maintaining stable executable instrumentation.}
Without the fixer, compilation failures increase dramatically. In the direct debugging setting, the number of compilation failures rises from zero to \textbf{94}. In the indirect debugging setting, this number increases from 2 to \textbf{106}. This large influx of compilation errors directly reduces the number of analyzable executions. Consequently, it leads to a lower recall rate. For instance, the recall drops to 0.366 compared to 0.511 in the full direct debugging setup. These results confirm that automatically repairing compilation errors introduced during logging insertion is critical. This repair ensures that the system can reliably collect actual runtime feedback.



\textbf{The iterative refinement mechanism enhances diagnostic utility and acts synergistically with compilation repair.}
Removing the refinement loop reduces the framework to a traditional static logging generator. Under the direct debugging setting, the detection F1 score drops from \textbf{0.520} to 0.388, and the total count of correctly detected bugs decreases from 159 to 118. The indirect debugging scenario shows a similar trend, with the F1 score decreasing from 0.408 to 0.364. These results confirm that refinement driven by runtime feedback incrementally captures execution information relevant to the underlying failures. Ultimately, these ablation findings highlight a complementary system design. While the repair module ensures execution stability by preventing instrumentation failures, the refinement mechanism leverages the resulting data to improve log informativeness. This synergy enables \tool to consistently generate logs that are both executable and diagnostically useful.

\textbf{Runtime feedback provides gains beyond multi-round source-based refinement.}
When runtime feedback is disabled but refinement remains available, the direct debugging F1 score decreases from 0.520 to 0.461, and successful repairs decrease from 97 to 76. In the indirect setting, F1 decreases from 0.408 to 0.368. This variant also generates more logs in the indirect setting, averaging 12.50 logs per caller, but still performs worse than full \tool. These results indicate that the benefit does not come merely from allowing more rounds or more statements. Runtime execution evidence helps the critic identify concrete observability gaps and helps the refiner add logs that explain the actual failing behavior.

\begin{figure*}[t]
\centering
\includegraphics[width=\textwidth]{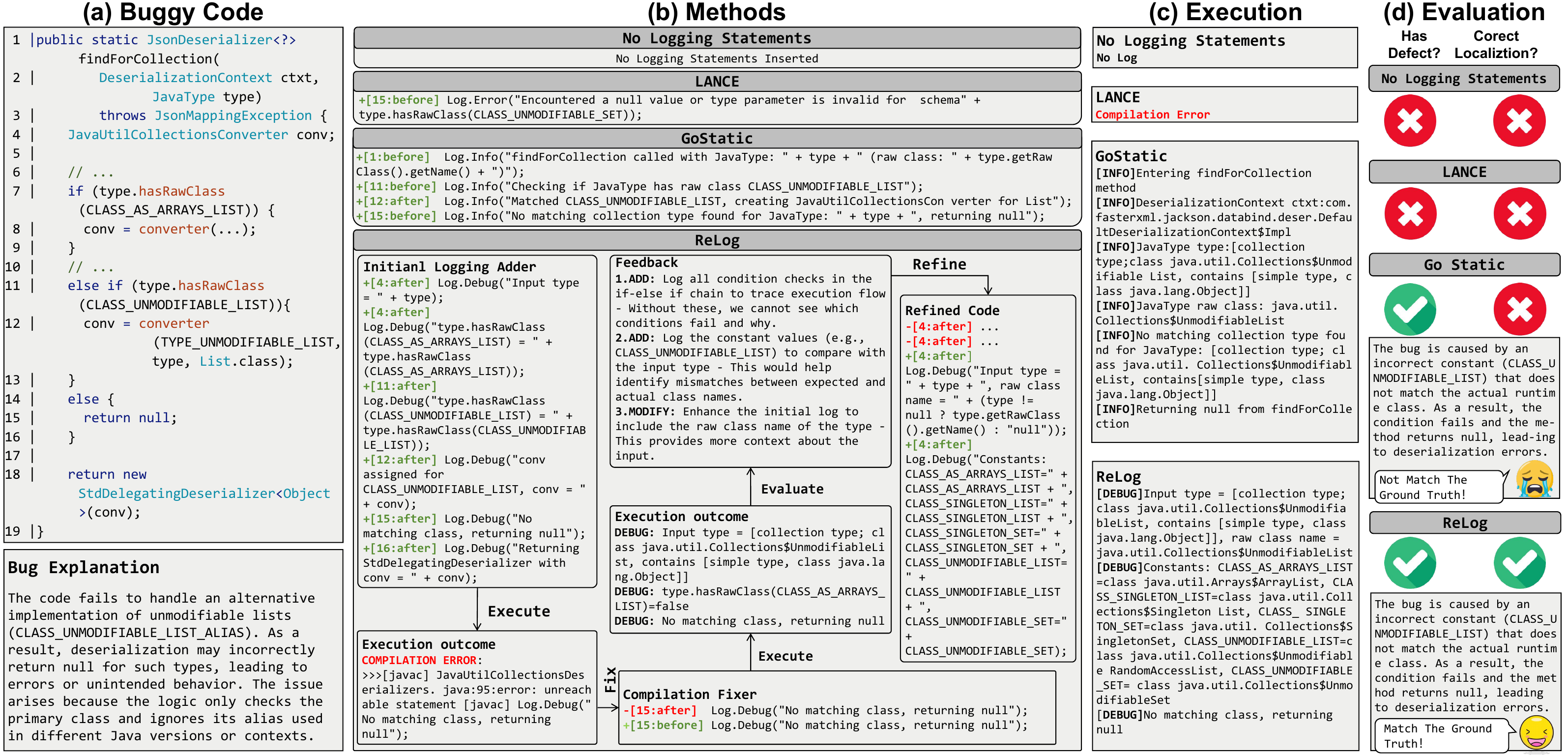}
\Description{A case study comparing ReLog and baseline logging methods on a representative defect, including inserted logs and final defect localization outcomes.}
\caption{A comparative case study of logging generation approaches on a representative defect. The figure illustrates the inserted logging statements, the collected runtime log, and the final defect localization outcomes for \tool and the evaluated static baselines.}
\label{fig:case_study}
\end{figure*}

\greybox{
\textbf{RQ3 Summary:}
The debugging effectiveness of \tool stems from the synergy between compilation repair, runtime feedback, and iterative refinement. The repair module ensures execution stability, runtime feedback exposes concrete observability gaps, and the refinement mechanism uses this evidence to progressively enhance logging quality.
}

\subsection*{Case Study}

To demonstrate how \tool enhances downstream debugging in practice, we conduct a qualitative case study on a representative defect from our benchmark. We compare the logging statements and the resulting execution traces produced by \tool against baseline approaches, specifically GoStatic and LANCE. This detailed examination reveals how our iterative generation paradigm successfully provides actionable evidence for automated fault localization compared to traditional static methods.

As illustrated in Figure~\ref{fig:case_study}, evaluating the original code without any logging statements causes the debugging agent to miss the defect entirely. Static generation approaches struggle to improve this outcome. For instance, LANCE introduces a compilation error during log insertion, which prevents execution and directly leads to a detection failure. While GoStatic successfully compiles the code and generates execution traces, these logs only record the method entry and the final unexpected return value. Because they fail to expose why the conditional logic bypasses the target branch, the agent detects an anomaly but incorrectly explains the root cause.

In contrast, \tool systematically constructs a complete chain of diagnostic evidence through compilation repair and iterative refinement. Initially, the framework places a log immediately after a return statement. The compilation repair module detects this unreachable code error and automatically repositions the statement before the return, ensuring that the instrumented program can execute. The first collected trace shows that the conditional check fails and the method returns a null value. Determining that this lacks sufficient context, the log sufficiency evaluator suggests logging the exact values of the class constants for a direct comparison. The refiner then augments the logging code to print both the input raw class name and the target constant values. The subsequent execution clearly exposes the hidden type mismatch. Armed with this precise structural evidence, the downstream agent correctly identifies the missing alias check and accurately explains the defect.

\section{Related Work}
\label{sec:related}
\subsection{Logging Statements Generation}
Many existing research explores the automatic generation of logging statements to assist developers in collecting runtime information software systems. These approaches typically address three key challenge in logging design. They determine the optimal locations for logging, the appropriate severity levels, and the specific information to record in the messages. Early studies focused on predicting logging locations by training machine learning models on source code features~\cite{zhu_learning_2015, li_where_2020}. Subsequent research extended this direction. Researchers applied deep learning techniques to recommend suitable logging levels based on the surrounding code context~\cite{li_deeplv_2021,heng2025benchmarking}. More recent approaches leverage neural sequence generation and LLMs. These advanced methods generate complete logging statements that include both the textual message content and the relevant code variables~\cite{mastropaolo_UsingDeepLearning_2022, mastropaolo_LogStatementsGeneration_2023, ding_logentext_2022, ding_logentext-plus_2023, xu_UniLogAutomaticLogging_2024, xie_FastLogEndtoEndMethod_2024, li2024exploring, li_go_2024, zhong2025beyond, zhong_log_updater, zhong_EndtoEndAutomatedLogging_2025}.


Despite recent advances, existing techniques generate logging statements through a static single pass over source code while ignoring program execution behavior. Furthermore, these methods evaluate generation quality using text similarity and position prediction rather than measuring debugging utility. In contrast, our work introduces an iterative framework that refines logging statements using runtime feedback and evaluates its effectiveness through downstream debugging tasks.

\subsection{Logging Statements Practice}
Some researchers investigates logging practices and defects related to logging in production software systems. Prior studies analyze logging usage patterns by mining version histories and issue reports. 
For instance, researchers characterize common logging practices and recurring challenges. These challenges include deciding what information to record and structuring the textual messages~\cite{chen_characterizing_2017}. Other studies examine specific logging issues in detail. They identify temporal inconsistencies between source code and logs~\cite{ding_temporal_2023}. They also detect duplicated logging statements~\cite{li_dlfinder_2019} and highlight readability problems in logging messages~\cite{li_are_2023}. Together, these empirical studies provide valuable insights into the design of logging statements and expose common pitfalls in practical software development and maintenance workflows.

While prior research primarily assists human developers through guidelines and pattern mining, the increasing reliance on LLMs demands logging strategies optimized for automated maintenance. Consequently, our work shifts the focus from human readability to generation tailored specifically for downstream software engineering tasks such as automated debugging.

\section{Threats to Validity}
\label{sec:threat}

\phead{Internal Validity.}
Threats to internal validity concern factors within our experiment that could influence the evaluation results. A primary threat is the dependency on structurally valid input code. Because the iterative refinement mechanism of \tool relies on dynamic execution, compilation errors in languages like Java block runtime signal extraction and could skew the performance assessment. To mitigate this, we evaluated our approach on a curated Java dataset where the input codebases are inherently executable, thereby isolating the effectiveness of our logging generation from basic syntax resolution.


\phead{External Validity.}
Threats to external validity concern the generalizability of our findings. A key threat is the applicability of \tool to diverse execution environments. Since iterative refinement introduces runtime overhead, its direct use in latency-sensitive production systems is limited. We mitigate this by scoping \tool to offline debugging workflows, such as CI/CD failure diagnosis and local reproduction, where developers can invoke the framework after failures occur. Another threat is that our benchmark is based on Java defects from Defects4J. Although it provides reproducible failures and ground-truth fixes, future work should validate \tool on more languages, projects, and bug types. Finally, we focus on automated debugging as the downstream task, as defect diagnosis is a critical use case for logging and offers a rigorous benchmark for log quality. Extending \tool to anomaly detection, performance monitoring, readability assessment, or runtime-overhead optimization remains future work and can be supported by adapting the critic rubric.

\section{Conclusion}
\label{sec:conclusion}
In this paper, we introduce \tool to overcome the limitations of static logging statement generation. Instead of relying solely on static code structures, \tool iteratively refines logging statements using LLMs guided by execution feedback. Extensive evaluations on direct and indirect debugging tasks show that this dynamic approach consistently outperforms existing baselines, generalizes across upstream and downstream LLMs, and benefits from runtime feedback beyond multi-round source-based refinement. By resolving compilation errors and optimizing log content through runtime observations, \tool generates informative execution traces even when the faulty source code is inaccessible to the downstream agent. Overall, this work establishes a new paradigm for automated logging generation, showing that prioritizing diagnostic utility over static text matching better bridges generated logs and downstream automated debugging.


\section{Data Availability}

To facilitate reproducibility and future research, the complete replication package for \tool is available~\cite{replication_package}.
Researchers can access all necessary scripts and instructions to reproduce the findings presented in this paper.

\section{Acknowledgments}
This research was supported by the National Natural Science Foundation of China (No. 62502409).

\bibliographystyle{ACM-Reference-Format}
\bibliography{main}

\end{document}